# The Resolution of the Diamond Problem After 200 Years:
## Phonon, Roton and Magnon Induced Spin-Orbital Dynamics,
## Subshell Rehybridization and Shell Rotation for the Little Effect

### Reginald B. Little[*] and Joseph Roache

### Florida A&M University

### Florida State University

### National High Magnetic Field Laboratory

### Tallahassee, Florida 32306

**Abstract:**

The problem of the physicochemical synthesis of diamond spans more than 200 years, involving many giants of science. Many technologies have been discovered, realized and used to resolve this diamond problem. Here the origin, definition and cause of the diamond problem are presented. The Resolution of the diamond problem is then discussed on the basis of the Little Effect, involving novel roton-phonon driven (antisymmetrical) multi-spin induced orbital orientation, subshell rehybridization and valence shell rotation of radical complexes in quantum fluids under magnetization across thermal, pressure, compositional, and spinor gradients in both space and time. Some experimental evidence of this magnetic quantum Resolution is briefly reviewed and integrated with this recent fruitful discovery. Furthermore, the implications of the Little Effect in comparison to the Woodward-Hoffman Rule are considered. The distinction of the Little Effect from the prior radical pair effect is clarified. The better compatibility of radicals, dangling bonds and magnetism with the diamond lattice relative to the graphitic lattice is discussed. Finally, these novel physicochemical phenomena for the Little Effect are compared with the natural diamond genesis.

* corresponding author; email : redge_little@yahoo.com

**The Diamond Problem:**

A long history of giants of science has defined and contributed to the solution of the diamond problem. The list includes Newton, Boyle, Lavoisier, Guillton, Clouet, Karazin, Despretz, Hannay, Moisson, Roozeboom, Jessup, Rossini, Tammann, Parson, Einstein, Leipunski, Bridgman, Berman, Simon, Hershey, von Platen, Lundblad, Hall, Bundy, Strong, Wentorf, Cannon, Eversole, Deryagin, Angus, Fedoseev, Setaka, Matsumoto, Yugo, Linarres, Hemley, Sumiya, and Little. On the basis of these investigators, many technologies have been considered for resolving the diamond problem. These technologies include electric arcs; metal solvents; metal catalysts; electric ovens; electric resistive heaters; anvil vices and presses; electron beams; atomic beams; x-rays, alpha, and beta irradiators; exploding media; rapid lasing and liquid nitrogen quencher; chemical vapor deposition (CVD); hydrogenous plasma and microwave apparatuses; hot filaments; flames; and **recently strong magnets**. Some of these technologies have resulted in partial successes by the direct method, the indirect method and the H plasma metastable method of forming diamond. The complete Resolution here considers all three of these methods along with natural diamond formation and demonstrates complete Resolution by external magnetization (using external magnetization in conjunction with these three methods and considering intrinsic magnetism in mantle Kimberlite).

**Magnetic Quantum Description:**

On this basis, the synthesis and understanding of diamond have developed along side developments in chemistry, physics and engineering during the last 300 years. In particular, the development of quantum theory in the last century and its detailed applications to matter are realized and put forth in this Resolution for a complete solution to the diamond problem, thereby providing a giant leap to solve the last piece of the puzzle of the diamond problem. Older strategies considered squeezing bulk atomic densities from the graphitic to the diamond allotrope on the basis of classic continuum of atomic densities. However such allotropic carbon transformations as determined here involve electronic alterations, which lack continuum of states. Therefore on the basis of low density of atomic states the transformation involves the dilemma of high temperature, high volume and low pressure conditions to break ("cut off") the graphitic spring and low temperature, low volume, high pressure contrary conditions to fix ("cut on") the diamond spring. Such contrary conditions contribute to the large activation barrier and the prior partial success by the gradients in these reaction conditions. The puzzle is Resolved here by determining and manipulating the nonclassical and high spin importance of the intermediary phases (during the direct; the indirect; and H plasma processes) associated with the pertinent activated reaction trajectories for diamond formation by realizing and controlling the high spin, magnetic, quantum, and fluidic natures of these intermediary phases during these diamond forming processes. This nonclassical nature of the diamond forming reaction trajectories follows from the low density of electronic states of carbon atoms and lack of inner core p subshell such that multi-atom quanta interactions are required for diamond symmetry and such large quanta and specific momenta of multi-atoms persist beyond atomic and molecular into mesoscopic and into macro-dimensions. Other $2^{nd}$ row elements may also suffer such fixation dilemma. 3d and 4f elements have less dramatic fixation difficulties and may provide catalytic effects for $2^{nd}$ row elements based on interactions providing symmetrical and higher electronic densities for necessary fruitful bond alterations. However, such low densities of states are characteristic of Bose-Einstein statistics for the quantum fluidic intermediates. Here higher densities of electronic states are realized by fermionic quanta fluidic intermediates for different Fermi-Dirac statistical dynamics. Therefore, the nonclassical reaction trajectories and Bose-Einstein statistics result in the large quanta and large activation energies. But the Resolution put forth here by magnetization and many body interactions and consequent Fermi Dirac statistics result in smaller quanta and lower activation barriers. Such fermionic, magnetic, quantum, fluidic conditions and media provide chemical energy, chemical volume (permittivity-permeability) and chemical pressures to lower thermo-mechanical volume, and pressure requirements of activation for forming diamond.

The Resolution determines the formation of these magnetic, quantum fluids for all processes that form diamond: the direct (liquid carbon), the indirect (liquid ferrometals), and the metastable (atomic H-plasma) processes. The synthesis conditions of higher pressures, higher temperatures and thermal plasma induce intrinsic magnetic properties in the growth media for fermionic statistics. These intermediary magnetic fluids behave nonclassically on microscopic and collective scales and in different ways relative to the more classical fluids like air and water.

**The Complete Resolution:**

Here the Resolution of the Diamond Problem is determined based on further taking advantage (via external magnetization) of these intrinsic magnetic nonclassical aspects of the carbonaceous (C), hydrogenous (H), and ferrometal (M) quantum fluids that define and lower the reaction pathways that form diamond by these various methods. Intrinsic magnetic aspects of such C, M, and H quantum fluids are identified as important to nucleate and grow diamond. In particular, magnetic characteristics of the C, M, and H fluidic solvents and magnetic solute complexes ($CC_4 \bullet C_x$), ($CC_4 \bullet M_x$) and ($CC_4 \bullet H_x$) are determined as important states along the reaction trajectories causing fruitful and unique dynamics of carbonaceous intermediates for $sp^3$ carbon formation, stabilization, orientation, organization, correlation, and knitting into diamond. See Figure 1. Such diamond-forming magnetic quantum fluids antisymmetrically prevent the collapse of intermediates to amorphous carbon and graphitic undesirables. The electronic natures (half-filled subshells) of the atoms of such fluids contribute to such fruitful magnetic quantum fluidic intermediates along the reaction trajectories. The antisymmetry further allows accumulation of dense, loosely bound radical media. The diamond forming conditions of the various techniques determine gradients wherein these radicals form and exist in hotter portions of the gradients and accumulate in cooler portions of the gradients. Across the gradients from hotter to cooler the dense radicals magnetically organize their orbitals, subshells and shells. The motions and magnetic interactions between the radicals lead to electronic rehybridizations. The C, M and H quanta fluidic solvent and the various carbonaceous complex solutes are reactants and products of many radical chemical reactions for novel chemical dynamics. See Figures 2 and 3. These diamond forming magnetic aspects and dynamics of the resulting quantum solutions involve novel phonon and roton driven (antisymmetrical) multi-spin induced orbital dynamics, subshell rehybridizations and valence shell rotations of central C and M atoms of these complexes by dense phonons, rotons and magnons within the quantum fluids (**known as the Little Effect[1]**) (See Figure 4.), which determine important transformations, complexations, transport, orientation, organization, rotations and release dynamics of carbonaceous intermediates that determine the temperature (pressure, compositional and spin) gradient driven reaction trajectories during diamond formation by the various techniques of the direct, indirect and H-plasma CVD processes. See Figure 1. Within the cooler regions of the gradients, the phonon-roton induced electronic subshell swirl and shell rotations slow, the radical complexes interact more strongly, and covalent bonds become locked into $sp^3$ configurations. Such dynamics of C, M and/or H radicals occur across gradients within magnetic domains. Many domains may exhibit different magnetic orientations for consequent polycrystalline diamond formation. External magnetization organizes many domains of these intrinsic magnetic effects on macroscale for larger single crystal diamond formation. The use of external magnetic field is put forth here as a new useful technology for enhancing these intrinsic magnetic effects and ordering, organizing and correlating magnetic complexes over larger space in these older synthetic techniques for even faster, larger, and better diamond synthesis.

The quantum fluidic nature of the diamond forming intermediates is consistent with recent fluids reviewed by Bigelow [2]. Bigelow [2] determines that spin and collisonal aspects of atomic bosonically interacting radicals can allow Bose Einstein condensation of atomic radicals. Little invokes similar effects but for Fermi-Dirac statistics for the diamond forming media for paramagnetic or possibly ferromagnetic ordering of radicals for these novel catalytic effects for fixing carbon atoms into $sp^3$ bonds under milder conditions. Many investigators have observed that increasing both temperature and pressure allow ferromagnetism to persist beyond the Curie temperature [3]. Similar to Bigelow [2] collisions and higher pressures contribute greater exchange and magnetic organization for the Little Effect [1]. The use of strong external magnetic field allows such novel magneto-catalysis of the Little Effect [1], involving conditions beyond the Curie temperature for phenomena undescribed by the Hedvall Effect [4]. It is important to consider that such extreme conditions contribute to the novel bond rearrangements according to the Little Effect [1] on the basis of the high spin and magnetism breaking the orbital symmetry for rehybridization and the frustration of the Woodward-Hoffman rule [4] under milder conditions. These novel dynamics of the Little Effect [1] for a new area of chemical dynamics involve dense antisymmetrical radical fermionic media and are much different from prior radical pair phenomena of Turro [6]and Buchachenko [8], which consider effects between only two radicals. It is important to note that the novel dynamics of the Little Effects [1] are not driven solely by magnetism. The bulk of the energy is derived from a furnace, a laser, hot wire or microwave. The magnetism only influences intermediates generated by these power sources.

**Evidence of Magnetic Diamond Formation:**

The predictions of the Little Effect and this Resolution are consistent with natural diamond formation and experimental data. Previously investigators have overlooked the magnetic aspects of Kimberlite pipes [8] and the magnetic influence on diamond formation in magmatic fluids in the earth's mantle and the magnetic protection of diamond during its volcanic emplacement to the earth's surface. Indeed, aeromagnetic survey is a very useful technology for locating diamond mines. Here is some evidence of what is to come by using external magnetism to ushers in a new era in diamond synthesis. Little [9] first discovered the accelerated formation of diamond in strong magnetic field (>15 Tesla) at atmospheric pressure by using Fe catalyst and carbon precursors. Wen [10] has subsequently observed similar effects of magnetic field on diamond formation although under the expansion pressure of sealed hot Fe pipes. Recently, Huang [11] suggest using magnetic field and microwave plasma on carbon polymer in order to enlarge a diamond seed. Druzhinin et al [12] used ultrastrong magnetic field (300 Tesla) for diamagnetic compression to crystallize diamond from graphite, but erroneously assumed the magnetic field only caused compression of graphite for diamond crystallization. Druzhinin did not explicitly use external magnetic field to directly influence the chemical dynamics. Druzhinin et al [11] did however note that the magnetic compression resulted in faster larger diamond crystals than comparable compression by traditional mechanical high pressure high temperature techniques. Druzhinin did not entertain, as done here in this Resolution, that the magnetic field itself contributes additional beneficial catalytic effects via radical intermediates for facilitating diamond nucleation and growth. Much weaker magnetic fields (> 1 Tesla) have been employed for over a decade to focus plasma during CVD; but not as put forth here wherein strong magnetization affects the antisymmetrical bond rearrangement at the growth interface as by the Little Effect [1]. Hiraki [13] used magnetic field in microwave plasma CVD but fields of much weaker strength than RB Little, serving simply to spread and densify the plasma for more uniform deposition by the Matsumoto style synthesis. Also, Wei [14] used that external magnetization to allow better internal coating of high aspect ratio tubes with diamond like carbon and SiC. R. B. Little is the first to predict, observe and discover that strong magnetization causes intrinsic nonclassical dynamics for enhancing bond rearrangement to diamond. Recently, Skvortsov [15] observed diamond formation in the strong magnetic monofields created by lasers.

In addition to supportive data of this Resolution by magnetic diamond formation, researchers have also observed that magnetization slows diamond decompositions during abrasion and grinding. Kuppuswamy [16] observed that external magnetic field affects and slows electrochemical etching and grinding of cemented carbides, diamond, SiC and $Al_2O_3$. Other researchers have reported the effect of magnetization in slowing decomposition of ferroabrasives [17]. Magnetism enhances metal removal but slows the damage to the diamond used to grind the metal. At only, 50-100 Gauss on 15% Na NO₃ removal rates of tungsten carbide was greater under the external magnetic field. On the basis of this Resolution, the magnetism slows the decomposition chemistry of diamond under the local high temperature and high pressure created by grinding processes such that graphitization is slowed and the diamond is protected. External magnetic field slows pi bonding, which is crucial for abrading diamond. ¶ bonds form due to impact and resulting high pressures and high temperatures during grinding but under external magnetic field and friction induced HPHT the diamond surface will not graphitize. The surface radicals formed during abrasion resist graphitization under the antisymmetry caused by the external magnetism. Graphitic structures are unstable in the radicals and magnetism.

**Magnetized Graphitic Instability:**

Experimental data of other investigators supports this graphitic instability in dense radical environments and under strong magnetic field. The diamond lattice is more stable than the graphitic lattice in the radical and magnetic environments due to the weaker coupling between the more localized dense $sp^3$ C-C bonds and lattice radicals. On the other hand, there is stronger unstable coupling between the more delocalized graphitic ring currents and formed lattice radicals. In support of this spinophillicity of diamond, ferromagnetic states in densely defective diamond have been observed [18]. Partridge et al observed ferromagnetic transfer into outer diamond coating on ferromagnetic cores when ferromagnetic metal rods are coated with diamond [19]. Putov et al. [20] observed that thermomagnetic treatment of steel produced excellent steel. There is a lot of evidence that the magnetization and high spin radical environments favor the spinophillic diamond lattice over the spinophobic graphitic lattice. Researchers have demonstrated the greater stability of spin in the diamond lattice and the formation of carbon onions in magnetic field by thermalizing nanodiamond in external magnetic field [21]. Whereas other researchers have demonstrated the weaker stability of spin in the graphitic lattice, by thermalizing nanodiamond in zero applied magnetic field to form turbostratic graphite and nanographite [22,23]. R. B. Little [9] observed that thermomagnetization (> 15 Tesla) of nanoparticles of Mo-Fe with flowing $CH_4$ and $H_2$ leads to nucleation of diamond at atmospheric pressure rather than the formation of CNT under similar conditions but in zero applied magnetic conditions. Yokomichi [24] observed this instability of graphitic structures and strong magnetism by the collapsed of growing CNT in strong external magnetic fields of 10 Tesla. B Wen [25] observed what he calls new diamond by thermomagnetized (10 Tesla) catalytic transformation of carbon black under higher pressure and Fe nanocatalysts to form this intermediary new diamond, thought to be intermediary between rhombohedral graphite and diamond. It is important to note that B. Wen et al.[10] needed background thermal expansion pressure at 10 Tesla to form diamond otherwise at 10 Tesla and atmospheric pressure they would have gotten the collapsed CNT of Yokomichi et al [24]. Therefore in the presence of high radical concentrations and strong magnetization, the carbon resists graphitization, favoring less delocalized bonds in the form of collapsed CNT, C-onions and nanodiamond. Such instability of radicals and magnetism with graphitic structures has been demonstrated to cause graphitic curvature for fullerenes, CNT and soot formation by the Comprehensive Mechanism of CNT Nucleation and Growth [26]. This impact of magnetic field and its creation of radicals for causing graphitic instability are consistent with the observations of many other researchers. Sun [27] observed that the application of weaker external magnetic field (<1 Tesla) during CNT synthesis eventually caused the CNT process to form amorphous carbon rods rather than CNT. Yokomichi et al. [24] observed that stronger magnetic fields up to 10 T caused greater $C_{70}/C_{60}$ formation during electric arc processes.

Raman [28] realized ring currents in graphite to explain the different magnetic susceptibilities of graphite and diamond. Here it is suggested in this Resolution that in CNT or fullerenes, these ring currents are oriented differently in space relative to planar graphite and this different orientations of ring currents in CNT and fullerenes cause their consequent internal ring---ring diamagnetic repulsions that intrinsically destabilize CNT (fullerenes) in strong magnetic field relative to nanodiamond. Such ring-ring interactions facilitate the spin transport in CNT [29]. Such effects of the ring currents in CNT are consistent with Kondo Effect [30] and Aharonov-Bohm Effect [31]. Haddon observed temperature and magnetic field dependence of susceptibility of various carbon structures [32]. Roche and Saito [33] observed that magnetism changes electronics of CNT at room temperature. Unlike graphitic structures, however many researchers have recently reported the stability, polarizing interaction and magnetism of magnetic impurity centers in diamond. High spin magnetic centers or multi-spin defects have been produced in nanodiamond for the emergence of its ferromagnetism [18,34]. The recent optically observed long lived triplet states [35] in diamond are further evidence of the greater affinity of diamond for spin intermediates and magnetic centers relative to graphitic allotropes. On the other hand, other researchers have demonstrated the greater incompatibility of radical media with the graphitic lattice. Enoki [36] demonstrates the strong coupling between such radical edge states and intrinsic graphitic ring currents. T. Enoki [37] recently demonstrated such high spin and magnetism in synthetic diamond frozen in the product during dynamical synthesis by explosion. Peng and coworker [38] observed such high spin and magnetism of fracture and cavities regions in black diamond formed by insufficiently high pressure. Just as broken C-C bonds in diamond form lattice C dangling bond, other lattice impurities may have dangling bonds, which the diamond lattice accommodates well relative to the graphitic lattice. Impurities like Ni, Co, Fe, N and H are accommodated by the diamond in a better way than the graphitic lattice. Defects are not as tolerated in graphitic structures due to the incompatible interactions of aromatic currents and the neighboring radicals. The pi cloud is super conducting and the radical is a magnet, so they do not mix in graphite.

**Conclusion:**

Diamond is a unique material beyond its extraordinary properties. The conditions for its growth are rather paradoxical. The simultaneous need for high volume, high temperature, low pressure conditions to activate its intermediates and for low volume, low temperature, high pressure conditions to organize and stabilize its condensation seem unrealistic over large space and short times. During the last 50 years limited resolutions of this diamond problem have been provided by the indirect high pressure high temperature methods, the direct super high pressure and high temperature method and the low pressure H vapor deposition. The magnetic quantum Resolution of this problem given here allows lower volume, lower temperature and atmospheric pressure by correlating radicals, atoms and other intermediates for greater formation, accumulation, organization and correlation of $sp^3$ carbon intermediates over larger space in shorter times for larger, faster single crystal diamond formation. On the basis of this magnetic quantum fluidic consideration of the reaction trajectories for diamond formation, the Little Effect determines novel electronic dynamics for clarifying the murkiness surrounding carbon integration into a diamond lattice for diamond nucleation and growth. This magnetic discovery for diamond synthesis has implications for other materials and processes as well. Use of external magnetization will present a new tool for controlling the reformation of hydrocarbons in conjunction with current high pressure and temperature, catalytic technology. Using superconducting magnets with such processes such as the Haber process will allow even less costly fixation of atmospheric nitrogen. The synthesis of boron compounds, singlet oxygen, and halogen oxides such as oxygen difluoride, dioxygen difluoride, and chlorine oxides may be facilitated in external magnetic field. Many borides form under high pressures and temperatures in the earth's sublayers. Halogen fluorides are important compounds for future rocket propellants, oxidizing and fluorinating agents. These halogen oxides cannot be produced directly from halogens and oxygen, so the use of electric arc allows limited formation. Magnetic methods may resolve the production of these boron compounds and halogen oxides just as the magnetization resolves diamond synthesis. Strong magnetic effects on some chemical processes present a new era of exploration with great impact on chemistry and physics.

**Acknowledgements**:

In gratitude to GOD.

For my three sons: Reginald Bernard Jr., Ryan Arthur and Christopher Michael.

Special thanks to Prof. Alan Marshall.

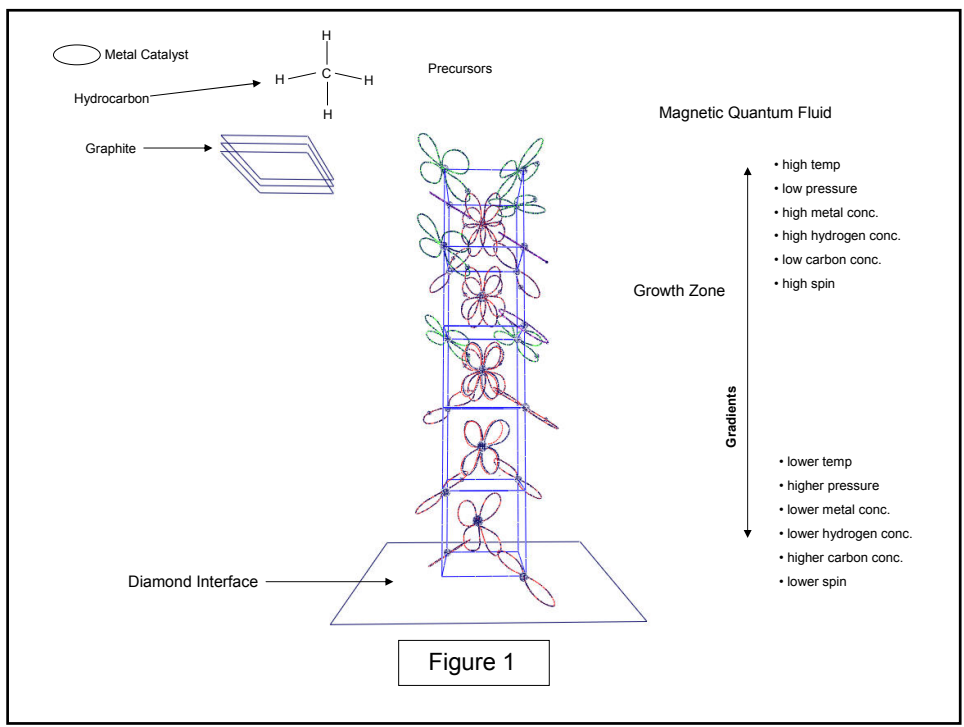

Figure 1

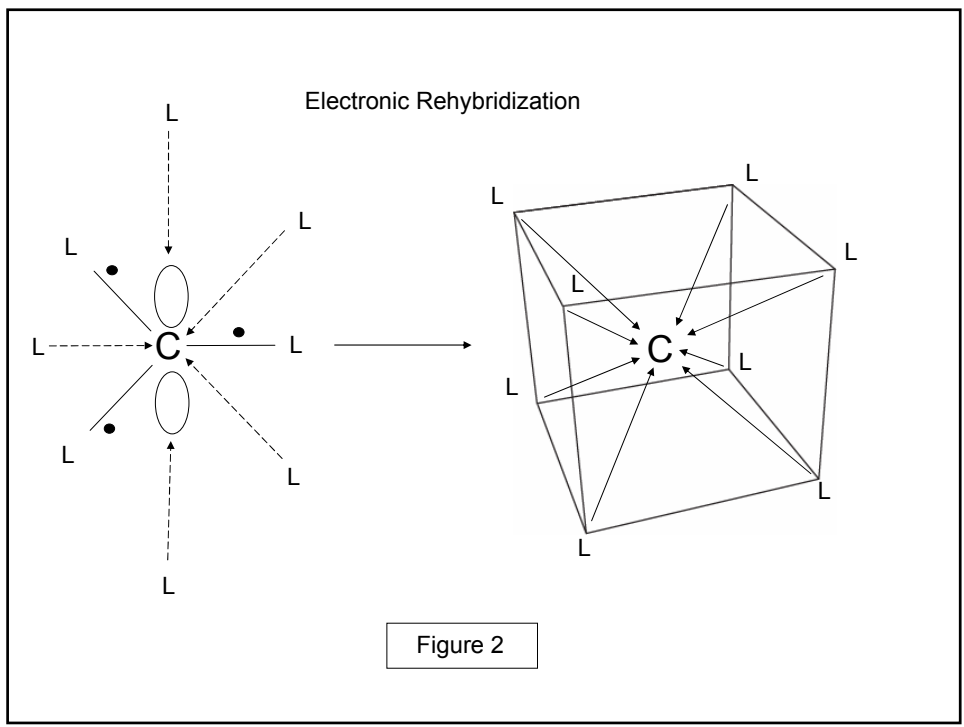

Figure 2

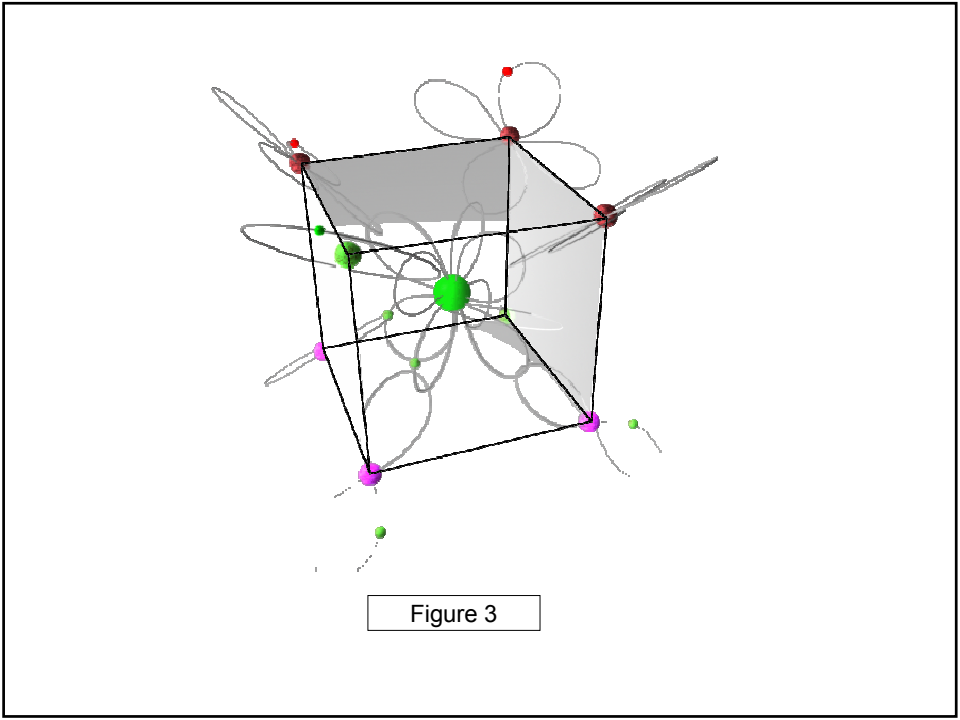

Figure 3

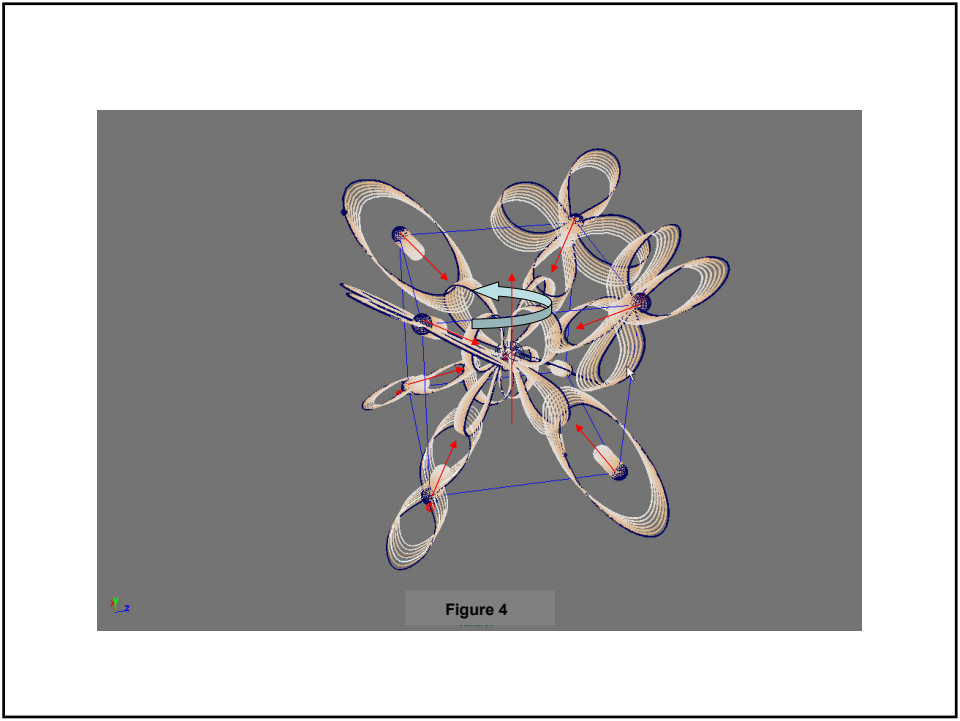

Figure 4